# Waves, Particles, and Quantized Transitions:
# A New Realistic Model of the Microworld

Alan M. Kadin*
July 25, 2011

**Abstract:**

A novel two-tiered organization of the microworld is presented, in which only the fundamental quantum fields of the standard model of particle physics (electrons, photons, quarks, etc.) are true quantum waves, exhibiting linear superposition. In contrast, confined quantum waves and their composites (such as nucleons and atoms) move collectively as particles following a classical Hamiltonian trajectory, as derived from the coherent phases of the component quantum waves. However, transitions between such quasi-classical trajectories are still subject to quantum transition rules of energy and momentum quantization (both linear and angular). Furthermore, there is no quantum decoherence, and no entanglement of multi-particle states. This provides a clear foundation for classical behavior, and avoids paradoxes of quantum measurement such as Schrödinger cat states. A synthesis of this type does not seem to have been previously examined. Can such a simple realistic representation really account for the known physics? This does require major reinterpretations of some established phenomena such as crystal diffraction, phonons, and superfluids.

*4 Mistflower Lane, Princeton Junction, NJ 08550
Email amkadin@alumni.princeton.edu

# I. Introduction: Wave-Particle Duality and Realism

The paradox of wave-particle duality has been embedded in the foundations of quantum mechanics since the very beginning.[1] How can a real object be both a localized particle and an extended wave at the same time? A wide range of physicists and philosophers have tried over many decades to make sense of this, without a clear resolution. The dominant trend has been make use of abstract mathematics according to standard procedures, without attempting to present a consistent, realistic physical picture. In this respect, quantum mechanics is anomalous; all other established physical theories have a clear physical picture that provides an accepted physical basis.

Given this long history, it would seem highly unlikely that yet another review of the same paradoxical physics could provide new insights. Indeed, one must go back to the beginning and reestablish a new set of more reliable and consistent foundations. In the present analysis, the assertion is made that only the fundamental quantum fields of the standard model of particle physics[2] (such as electrons and photons) are true quantum waves, and that composite entities (such as nucleons and atoms) are not. This flies in the face of the conventional belief that de Broglie waves are a universal aspect of all matter.[3] But back in the 1920's when quantum mechanics was first being developed, this distinction between fundamental and composite "particles" was not yet understood, so a universal model seemed quite appropriate.

Since then, the universal quantum wave paradigm has been used to explain a wide variety of physical phenomena, including, for example, neutron diffraction in crystals and quantized vibrations in molecules and solids. This may help to explain why a two-tiered quantum picture such as that proposed in the present analysis does not seem to have been seriously considered. This new picture needs to provide alternative explanations for these established phenomena.

Consider the neutron, for example. The measured scattering cross section proves that a neutron is a small particle ~ 1 fm in diameter.[4] But the conventional understanding of neutron diffraction[5] requires that a thermal neutron be a coherent extended wave much larger than its de Broglie wavelength $\lambda = h/mv$ ~ 100 pm. It is logically inconsistent to assert (as does the conventional interpretation) that a neutron is both a small particle and a wave that is $10^5$ times larger. The only way out of this paradox is to declare that only one of these pictures is real. It is argued below that the neutron particle picture is correct, and that there is no de Broglie wave for a neutron. Further, neutron diffraction does *not* require an extended coherent neutron wave, but rather reflects quantized momentum transfer of the ground-state crystal. Given the central role of wave diffraction in the history of physics, this is a rather radical assertion, but in fact a similar suggestion was made as far back as 1923,[6] and quantized momentum transfer in periodic structures was recently analyzed in more detail.[7,8]

A second type of paradox in conventional quantum theory deals with quantum measurement and the interface between quantum and classical phenomena. The conventional approach to quantum measurement requires a classical measuring instrument (or a classical observer) that introduces decoherence to a complex quantum wave,[9] but the basis for classical behavior in composites of quantum waves has never been quite clear. Furthermore, the conventional abstract formalism leads to complicated entanglement in composite states,[10] which in turn requires effects that are incompatible with local realism.

In contrast, the present analysis derives a classical Hamiltonian particle trajectory of a confined quantum wave from the Schrödinger equation of its internal components. In other words, a coherent localized quantum phase is responsible for classical behavior. However, all transitions between such quasi-classical states (including measuring instruments) are still quantized, and quantum effects are only negligible on the macroscopic scale if the separations of energy levels are small. This provides a more natural transition between the quantum and classical worlds, which is compatible with local realism.

In Section II, the evidence for wave and particle aspects of quantum systems is reviewed, with an eye toward clarifying the distinctions between electrons on the one hand and nucleons or atoms on the other hand. This is followed in Section III with a derivation of classical Hamiltonian trajectories from the Schrödinger equation for a simple model confined quantum wave. This concept of particle trajectories is applied in Section IV to diatomic molecules with internal degrees of freedom associated with rotation and vibration, where the basis for energy level quantization is assessed. This is extended in Section V to vibrational modes of crystals (i.e., phonons). An important implication is that crystal diffraction is a consequence of momentum quantization of the crystal, without requiring extended coherent wave effects in diffraction by neutrons or atoms. Section VI deals with the question of how quantum effects (including superfluid effects) can be extended to composite and macroscopic systems, without an explicit wave function of the composite object. Section VII briefly addresses the basis for quantization of the fundamental quantum fields. Further implications of this picture for quantum measurement and quantum information are discussed in Section VIII.

The picture presented here constitutes a consistent realistic picture of the microworld that does not seem to have been previously explored. While it is not (yet) a complete theory, it clearly identifies the necessary components of such a theory. This should encourage others to re-analyze other quantum paradoxes, to determine whether they might be resolved using a viewpoint similar to the one developed here.

## II. Evidence for waves and particles

In discussing wave-particle duality in quantum mechanics, it is important to clarify what we mean by each of these terms. A classical wave is a time-varying spatially extended field, a real distributed oscillation in real space, generally described by a linear wave equation, which can be analyzed in terms of wave components with ω and **k**, and is subject to linear superposition. Such a real wave may be modeled by a complex wave for mathematical convenience. A wave does not have a fixed size, but may be confined by boundaries to a region that is larger than half a wavelength. In this respect, a wave is like a gas in that it can expand to fill whatever box confines it. Such a confined wave forms a discrete (but infinite) set of standing wave modes (eigenstates), each with its own value of ω. This is the case, for example, with classical electromagnetic waves, where the amplitude of these wave modes (or their linear combinations) can take any value. A classical wave may carry distributed energy, momentum, and angular momentum, but these values are not quantized. Non-interacting waves can generally share space and pass through each other. There are scalar waves and vector waves; a vector wave may have polarization. Waves may be coherent or incoherent in space and time; coherent waves are associated with effects such as interference and diffraction.

This is distinguished from a true quantum wave, where not only are the mode frequencies quantized, but also their amplitudes are quantized. This gives rise to quantization of angular momentum (spin), linear momentum, and energy. This aspect is not explained by standard theory, but is briefly addressed below in Section VII. It is often asserted that a quantum wave is fundamentally a complex wave $\sim\exp(i\phi)$, rather than a real oscillation in real space, but this is really an artifact of the mathematical model rather than a fundamental aspect. In fact, the real oscillation frequency of a quantum wave is given by its full relativistic energy ($f=mc^2/h$), but this frequency is generally offset in the non-relativistic case. In terms of standing waves and superposition, a quantum wave behaves much like a classical wave. Transitions between waves of differing quantized amplitudes, of course, have a distinct quantum character.

In contrast to a wave with its infinite degrees of freedom, a classical particle follows a trajectory in space, associated with a classical Hamiltonian, with only three spatial degrees of freedom. It exhibits a fixed mass and internal structure, and its interactions follow conservation of energy, momentum, and angular momentum. One type of trajectory is a periodic oscillation or rotation, and the amplitude of this oscillation is unrestricted. One can also have a set of two or more interacting particles, the motion of which can (in many cases) be decoupled into collective modes that are largely independent. A particle generally has a defined size with a center of mass – point particles are mathematical idealizations which do not exist in nature.

In analogy to the quantum wave, let us also introduce the novel concept of a "quantum particle" which follows a classical trajectory, but which may also have a localized coherent phase, and the trajectory may have quantized amplitude. As described further in Section III, it is suggested that a confined quantum wave (such as in an atom or a nucleon) acts in its external motion as a quantum particle. This may be generalized to a quantum collective mode of two or more such quantum particles.

The distinction between quantum waves and classical particles may be illustrated by focusing on the quantum Hamiltonian $H_q$ and the classical Hamiltonian $H_c$, and how they are physically quite different, even though they have some formal similarities. The classical Hamiltonian $H_c(p,x)$ represents the total energy of a classical particle as a function of momentum p and position x (here one-dimensional for simplicity) and governs its trajectory x(t), given by the following pair of equations[11]

$$\frac{dx}{dt} = \frac{\partial H_c}{\partial p}, \qquad (1)$$

$$\frac{dp}{dt} = -\frac{\partial H_c}{\partial x}. \qquad (2)$$

For the simple case of a particle of mass m moving in a potential V(x), one has

$$H_c(p,x) = \frac{p^2}{2m} + V(x), \qquad (3)$$

which leads to the trajectory given by

$$\frac{dx}{dt} = v = p/m \qquad (4)$$

$$\frac{dp}{dt} = -\frac{\partial V}{\partial x}, \qquad (5)$$

which of course are exactly the equations given by Newton's law of motion F=ma for a force F=-∂V/∂x and acceleration a=dv/dt.

This should be contrasted with the quantum Hamiltonian operator $H_q$ for a quantum wave of mass m, which maps the momentum p to the functional operator -i$\hbar$ ∂/∂x (from the de Broglie relation p = $\hbar$k), yielding the relation

$$H_q = -\frac{\hbar^2}{2m}\frac{\partial^2}{\partial x^2} + V(x). \qquad (6)$$

The quantum Hamiltonian, in turn, forms the basis the Schrödinger equation[12] for the wavefunction Ψ(x,t)

$$H_q\Psi = i\hbar\, \partial\Psi/\partial t, \qquad (7)$$

using the energy operator (from E=$\hbar$ω) i$\hbar$ ∂/∂t. This is generally applied to the constant energy case for an appropriate quantized $E_n$ to yield

$$\Psi(x,t) = \exp(-iE_n t/\hbar)\, \psi(x), \qquad (8)$$

where $\psi(x)$ is the spatial part of the wave function given by the solution of the eigenvalue equation $H_q\psi = E_n\psi$, $|\psi|^2$ represents its distributed intensity, and $\int|\psi|^2 dx$ over the entire wave is typically normalized to unity. A distribution is not a generalized trajectory; rather, a trajectory can be derived from a distribution in the special case where all of the internal degrees of freedom are frozen. An example illustrating this is derived in Section III below.

Table I summarizes conventional evidence for wave nature and for particle nature in electrons, photons, neutrons, and atoms. The key point is that there is strong evidence for wave behavior in electrons and photons, and strong evidence for particle behavior in the external motion of nucleons and atoms (both indicated by bold italics in the table). In contrast, the evidence for particle behavior in electrons and photons tends to be more inferential, as does the evidence for wave behavior in nucleons and atoms (both indicated by parentheses in the table). The details are discussed in more detail below.

*Table I. Summary of Evidence for Waves and Particles*
*(Stronger evidence indicated by bold italics; weaker evidence by parentheses.)*

| Entity | Evidence for Waves | Evidence for Particles |
|---|---|---|
| Electron | ***Standing waves, Directional orbitals*** | (Quantized spin, mass, charge) |
| Photon | ***EM Waves*** | (Quantized spin, energy) |
| Atom | (Quantized vibrations in molecules) | ***Defined size in molecules & solids*** |
| Neutron | (Crystal diffraction) | ***Defined size in collisions & nucleus*** |

Standing waves in electronic orbitals constitute the clearest direct evidence of the wave nature of electrons. These occur in quantized orbitals in atoms, as well as in Bloch waves near the energy gap of crystalline solids. As a specific example that shows standing waves most clearly, consider the P orbital in an atom, with $L=\hbar$ quantized angular momentum corresponding to a 360° phase shift going around the nucleus (Fig. 1a). P orbitals in molecules and solids are generally standing waves (such as $P_x$) comprised of two oppositely directed traveling waves, forming rotational standing waves with a fixed angular node and antinode (Fig. 1b). This gives rise to the ubiquitous directional bonds in molecules and solids, which do not maintain rotational symmetry. Directional electron bonds in solids are strong direct evidence for electron quantum waves.

There are certainly quantized rotational states of molecules (in gases) which are conventionally attributed to rotating solutions of quantum wave equations. However, both nuclei and atomic orbitals are confined quantum waves with internal structure, but the rotational motion of these confined quantum waves is essentially classical, as indicated in Fig. 1c. In contrast to the electron rotational standing waves that are present in solids, molecular rotational states are never excited in solids, because the rotational standing waves, the molecular equivalents of $P_x$ orbitals, are not present in molecular motion. Indeed, if one can have quantization of angular momentum in quasi-classical molecular rotation, there is no need to assert atomic quantum waves; a quantum particle picture is more consistent.

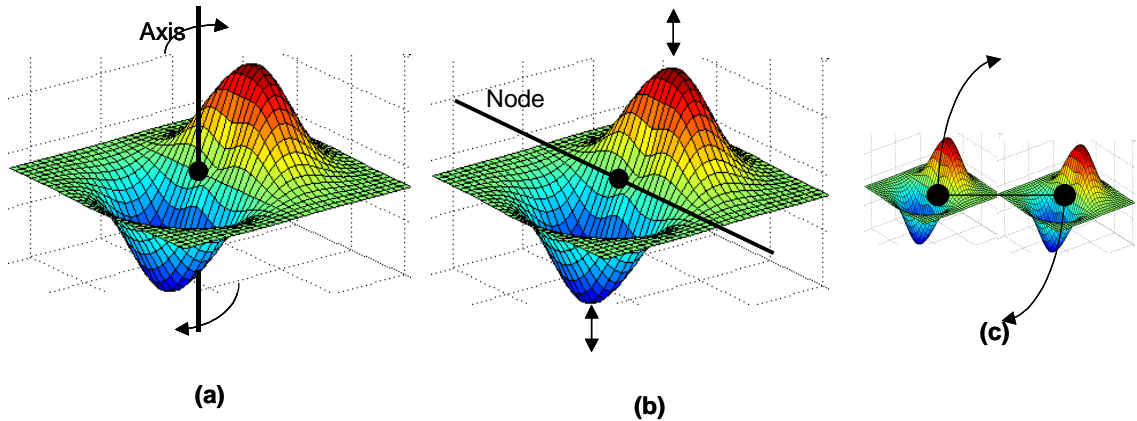

*Fig. 1. Quantized rotational states in electrons and molecules.*
  *(a) Rotating P-wave orbital with $\Delta\phi = 360°$ around nucleus.*
  *(b) Superposition of oppositely rotating P orbital components to give $P_x$ directional bond with fixed angular node.*
  *(c) Diatomic molecular rotation without apparent molecular wavefunction.*

Several other phenomena are also generally taken to prove the existence of quantum waves. Chief among these is crystal diffraction,[13] which is observed not only for electrons and photons, but also for neutrons and atoms and even molecules (Fig. 2). And certainly, a coherent extended wave with wave vector **k** impinging on a classical crystal can produce a coherent diffracted wave with $\Delta \mathbf{k} = \mathbf{G}$, where **G** is a reciprocal lattice vector (peaks in the spatial Fourier transform) of the crystal. However, one can obtain the same result for small particles (much smaller than the wavelength) impinging on the crystal, provided that the momentum transitions of a "quantized crystal" are restricted to changes $\Delta \mathbf{p} = \hbar \mathbf{G}$. This will be addressed further in Section V.

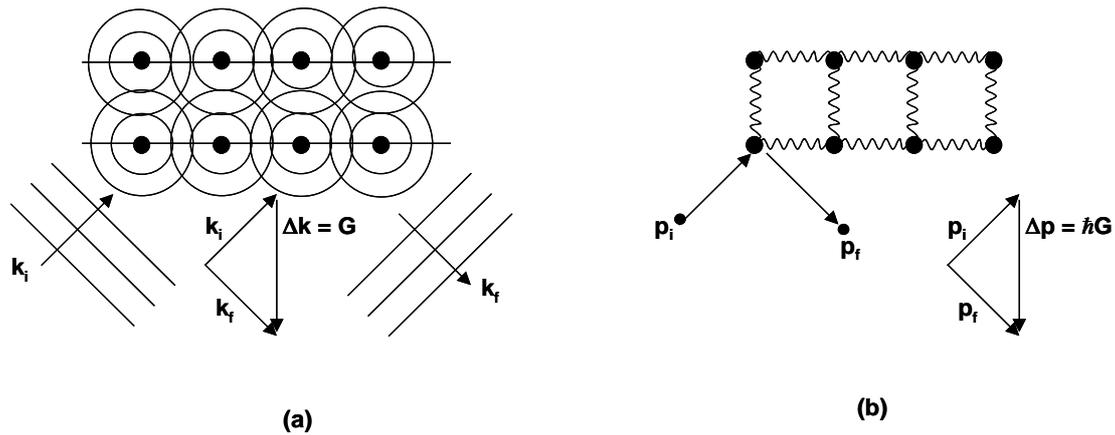

*Fig. 2. Crystal diffraction in wave and particle pictures.*
  *(a) Classical wave diffraction from periodic lattice with reciprocal lattice vector **G**.*
  *(b) Particle diffraction from quantized crystal with phonon having $\mathbf{p} = \hbar \mathbf{G}$.*

Another phenomenon that is also generally taken to prove the existence of quantum waves is quantization of energy. And certainly the quantized energies of the hydrogen atom exactly match the eigenstates of the Schrödinger equation, effectively proving that electrons are quantum waves. (The same argument can be made for muons, another fundamental quantum field, in hydrogenic atoms.) However, the situation is less clear for atomic motion in molecules and solids. These vibrational and rotational states are also quantized in energy, which is usually derived from solution of an appropriate Schrödinger equation. However, atoms (and nucleons, for that matter) have well-defined sizes and act like particles. It will be argued in Section IV that quantization of energy and angular momentum of photons can lead to quantization of energies in vibrating and oscillating modes of quantum *particles*, which in terms of their external motion are not really waves at all.

Furthermore, macroscopic quantum effects are seen in a variety of superfluid states of condensed matter: in superconductors, liquid helium, and alkali atoms (Bose-Einstein condensates). These effects are conventionally attributed to quantum waves from composite bosons. However, it is shown in Section VI that an alternative explanation in terms of primary electron wavefunctions may also account for these phenomena. There is no need to invoke a composite boson wavefunction, which in the present analysis does not exist.

It is sometimes asserted that an electron, and even a photon, is a point particle.[14] But these are clearly quantum waves, with no evidence of a point singularity. The primary argument for such a localized point particle is that this provides an apparent explanation for quantized mass and spin. For a distributed wave, the basis for such quantization is less obvious, and will be discussed further in Section VII.

There have also been interpretations of quantum mechanics that assert that a quantum entity is comprised of both a wave and a point particle. For example, within the conventional statistical interpretation, the Schrödinger equation describes probability waves, where the physical electron is a point somewhere within the envelope. Another less broadly accepted interpretation is the pilot wave picture,[15] due to de Broglie and Bohm, whereby the wave guides the trajectory of a point particle that is somewhere within the wave envelope. However, while these dualistic pictures may be mathematically consistent with the Schrödinger equation, they seem rather artificial.

A more natural picture would be one in which classical particle behavior evolves out of quantum wave behavior on the microscopic scale. A simple model that shows this is described in Section III below.

### III. Confined Waves and Particle Trajectories

In this section, we will show how classical particle trajectories can be derived from the motion of confined quantum waves. The intent is to explain the behavior of composite particles such as atoms or nucleons, but for simplicity let us first consider the motion of a one-dimensional electron wave function. In a constant potential V=0, the wave function takes the form of a plane wave

$$\Psi(x,t) = \exp(ik_0 x - i\omega_0 t), \tag{9}$$

where $E = \hbar\omega_0 = \hbar^2 k_0^2/2m$ and $p = \hbar k_0 = mv_0$, which is spread out over all space and is not normalizable, and cannot be associated with a "particle". One can construct a wave packet out of such plane wave components with a bandwidth $\delta k$ around a main component $k_0$, but such a wave packet necessarily includes components with a frequency bandwidth $\delta\omega$ and hence $\delta E$. The center of this wave packet follows a quasi-classical trajectory with a group velocity

$$v_g = \partial\omega/\partial k = \hbar k_0/m = v_0, \tag{10}$$

which does indeed correspond to a motion of a particle following the classical Hamiltonian equation (1) for $E=p^2/2m$. However, this wave packet is dispersive (since $v_g \propto k$ and $\delta k > 0$), spreading out as it propagates. Furthermore, the size of the wave packet ($\sim 1/\delta k$) is necessarily greater than the wavelength $\lambda = 2\pi/k_0$, generally $>>\lambda$ for a narrow $\delta k << k_0$. These are not the characteristics that one would associate with a classical particle.

Now consider a one-dimensional electron wave function confined between $x=0$ and $x=\ell$ by walls of infinite potential. The eigenstates take the form of sine waves based on standing waves with components $\exp(\pm k_n x)$, where $k_n \ell = n\pi$, for n=1, 2, ... so that

$$\Psi_n(x,t) = A \exp(-iE_n t/h)\sin(k_n x)$$
$$= (A/2i) \exp(-iE_n t/h) [\exp(ik_n x) - \exp(-ik_n x)] \tag{11}$$

This corresponds to traveling wave components with $\pm k_n$, with wavelength $\lambda_n = 2\pi/k_n = 2\ell/n$, and energies $E_n = \hbar\omega_n = \hbar^2 k_n^2/2m$. This has nodes at the two ends at $x=0$ and $x=\ell$, and also at intermediate points $x=\ell/j$, for j=1...n-1. While this is a particularly simple solution, confined wavefunctions generally include standing waves and nodes.

Now assume that this entire wavefunction (together with the box potential) is moving with a velocity $v_0$ to the right (positive x), or equivalently change to a reference frame moving with velocity $v_0$ to the left. This requires Doppler shifts of each of the plane wave components, one increasing in velocity, the other decreasing. The velocity shift corresponds to a wavevector shift of $k_0 = mv_0/\hbar$ for each of the plane wave components in Eq. (11), so that $k_\pm = k_n \pm k_0$. The energies for the two wave components are also modified accordingly:

$$E_\pm = \hbar\omega_\pm = \hbar^2 k_\pm^2/2m = (\hbar^2/2m)(k_n^2 + k_0^2 \pm 2k_n k_0) \quad (12)$$

Then the moving wavefunction becomes

$$\Psi_n(x,t) = (A/2i)[\exp(-i\omega_+ t + ik_+ x) - \exp(-i\omega_- t - ik_- x)]$$
$$= A\exp(-i\omega' t + ik_0 x)\sin[k_n(x - v_0 t)]$$
$$= A\exp(i\phi)\sin[k_n(x - v_0 t)], \quad (13)$$

where

$$E' = \hbar\omega' = (\hbar^2/2m)(k_n^2 + k_0^2) = E_n + \hbar^2 k_0^2/2m, \quad (14)$$

and where Eq. (13) applies only inside the moving box; $\Psi_n = 0$ outside the box defined by $(x - v_0 t)$ in the interval $[0, \ell]$. This new wave function represents a phase factor $\exp(i\phi)$ modulated by an envelope function. The envelope function with all of its nodes is moving with velocity $v_0$ to the right. Furthermore, the group velocity of the phase factor also corresponds to $v_0$:

$$v_g = \partial\omega'/\partial k_0 = \hbar k_0/m = p_0/m = v_0. \quad (15)$$

This acts like a particle of mass m, having internal energy $E_n$, moving with velocity $v_0$ (from classical Hamiltonian equation (1)). However, this is not a wave packet, in that the size of the envelope L is fixed and independent of the effective wavelength (the distance over which the traveling phase factor $\phi$ changes by $2\pi$) $\lambda = 2\pi/k_0 = h/mv_0$, so that one can have $\ell \ll \lambda$ (as in Fig. 3). The internode distances can be even smaller. Of course, the internal structure of the confined quantum wave does contain short-wavelength components, but that is distinct from the wave components corresponding to $\phi$ and $v_0$. From the external point of view, this box is a real object with a classical trajectory.

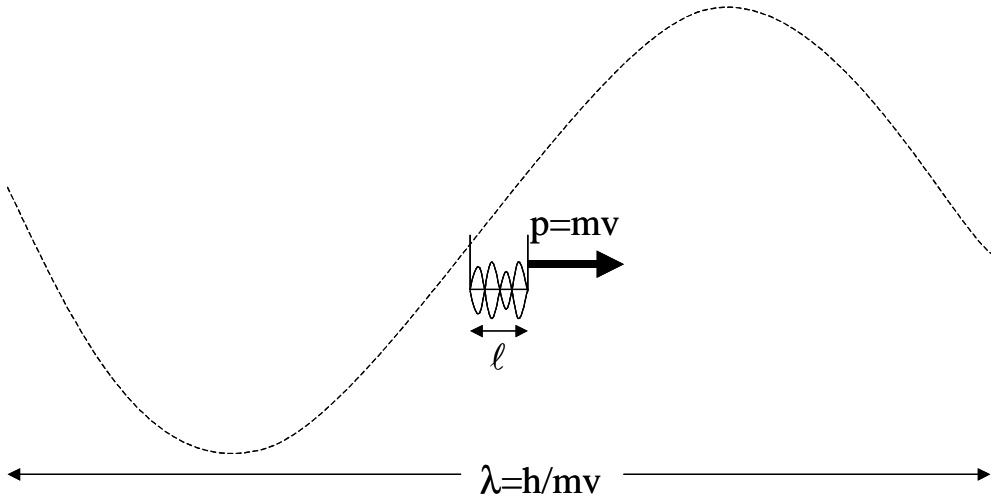

*Fig. 3. Trajectory of confined quantum wave (with internal standing-wave structure), moving with velocity v, with effective wavelength (shown in dashed line) $\lambda = h/mv \gg \ell$.*

One can now go further and add a potential V(x) that changes very slowly on the scale of $\ell$, in addition to the "box" potential. The relevant portion of the potential, of course, is that inside the box, which is essentially constant and will not perturb the bound solutions. The energies in Eqs. (12) and (14) are modified by simply adding V(x):

$$E_\pm = \hbar\omega_\pm = (\hbar^2/2m)(k_n^2 + k_0^2 \pm 2k_n k_0) + V(x), \qquad (16)$$

$$E' = \hbar\omega' = (\hbar^2/2m)(k_n^2 + k_0^2) + V(x), \qquad (17)$$

where the wavefunction takes exactly the same form as in Eq. (13). Since the total energy E' is a constant of motion in quantum mechanics, and it is assumed that $k_n$ also remains fixed, an increase in V(x) must produce a corresponding decrease in the external kinetic energy $\hbar^2 k_0^2/2m$. One can express this in terms of the change in $\omega'$ along the trajectory x(t):

$$\frac{d\omega'}{dt} = 0 = \frac{\partial \omega'}{\partial k_0}\frac{dk_0}{dt} + \frac{\partial \omega'}{\partial x}\frac{dx}{dt} = v_0 \frac{dk_0}{dt} + \frac{1}{\hbar}\frac{\partial V(x)}{\partial x} v_0. \qquad (18)$$

This yields the relation

$$\hbar \frac{dk_0}{dt} = -\frac{\partial V(x)}{\partial x} = \frac{dp}{dt} = -\frac{\partial H_c}{\partial x}, \qquad (19)$$

if one defines the classical Hamiltonian as in Eq. (3). So the motion of this confined quantum wave follows a trajectory given by both Eqs. (1) and (2) of the classical Hamiltonian formalism.

Thus, this system is effectively a classical particle with a trajectory x(t) (which may represent the center of mass of the object), which also has a phase $\phi(x,t)$ given by

$$\phi(x,t) = (p_0 x - E't)/\hbar \qquad (20)$$

Given the usual classical Hamiltonian relating E to p, the quantum equations have reproduced the classical trajectory, as long as the internal structure remains constant. Unlike the free electron, there is no quantum uncertainty or wave dispersion associated with this trajectory. There is only the single classical degree of freedom associated with collective motion. In the limit that $\ell \ll \lambda$, both the particle position x and its momentum p can be narrowly defined, in apparent violation of the Heisenberg uncertainty principle.[16] But Eq. (20) is not the phase of an extended wave; rather, it is a localized oscillator.

A composite of two or more confined electrons or other quantum waves bound together would each exhibit their own phase factors, and would follow essentially the same trajectory. This would apply, for example, to an atom or ion moving in a potential. The conventional assertion that the entire atom creates its own de Broglie wave associated with the total momentum is inconsistent with this physical picture. The same considerations should also hold for a nucleon (such as a neutron), which is fundamentally a set of three bound, confined quark wavefunctions.[17] Since conventional matter is composed of nuclei and atoms, this picture should apply quite generally. We will return to this later in the context of neutron diffraction.

It is important to note that the classical trajectory of this quantum particle is compatible with the presence of coherent oscillations of a quantum phase factor. In fact, these coherent oscillations are essential for obtaining classical behavior. This contrasts sharply with the conventional view that quantum coherence needs to be lost to recover classical behavior.[9] Furthermore, these oscillations provide a real spacetime basis for the classical total energy and momentum. The total energy of a particular quantum particle is given by the quantum oscillation frequency $f = E'/h$, where $E'$ is properly the full relativistic energy in the laboratory frame, including kinetic and potential energy terms. For a composite, the total energy is the sum of those of each of the fundamental components.

This picture also seems to violate Bohr's Correspondence Principle,[18] by which classical behavior is recovered only in the limit of large quantum numbers. This analysis derives a classical trajectory from a wave equation for a confined quantum wave, where the internal structure may be in any quantum state. Moreover, some classical aspects may never be recovered for any quantum number. For example, if one has a harmonic oscillator potential, then a classical oscillator can have any oscillatory amplitude. On the other hand, if this potential is based on electromagnetic interactions, and electromagnetic waves (i.e., photons) are quantized, this quantization may restrict the allowable amplitude of the "quantum particle", as indicated in the analysis below in Section IV.

**IV.   Diatomic Molecules and Quantized Transitions**

This quasi-classical trajectory seems quite different from the conventional picture, in which each object composed of quantum waves exhibits its own quantum interference and uncertainty. If one can directly derive classical mechanics from confined quantum waves, why is this conventional picture universally believed? The main reason seems to be that there are a variety of quantum effects involving composite particles. Several of these are examined here, including molecular vibration and rotation.

Consider first the case of vibration of a diatomic molecule with masses $M_1$ and $M_2$ and positions $\mathbf{r_1}$ and $\mathbf{r_2}$.[19] Each atom consists of the atomic nucleus plus all of the bound electrons. The conventional approach is to first generate the classical Hamiltonian corresponding to the internal reduced mass $\mu = M_1 M_2/(M_1+M_2)$ and the distance between the atoms $R=|\mathbf{r_1}-\mathbf{r_2}|$, where $R_0$ is the equilibrium separation:

$$H_c = p^2/2\mu + K(R-R_0)^2/2 \qquad (21)$$

using the standard quadratic approximation of the interatomic potential. Classically, this yields simple harmonic motion of the form

$$R(t) = R_0 + A\cos(\omega_0 t), \qquad (22)$$

where $\omega_0 = (K/\mu)^{0.5}$ and A is an arbitrary amplitude. The resulting energy is $E = A^2 K/2$, which can also have any value. Both atoms would contribute to this, oscillating in opposite directions at the same frequency.

In conventional quantum mechanics, however, $H_c$ is mapped onto the quantum Hamiltonian for a distributed wavefunction (as opposed to an oscillating trajectory), just as if it were an electron in a potential, but with a reduced atomic mass $\mu$. This Schrödinger equation is solved to generate a discrete set of eigenstates, which consist essentially of standing waves with nodes. The allowed energies are $E_n = (n+1/2)\hbar\omega_0$ for $n = 0, 1, ....$ And indeed, infrared spectroscopy confirms transition energies $\hbar\omega_0$, and the specific heat of gases and solids also requires the discrete energy levels.

These distributed standing quantum waves are physically quite different from oscillating trajectories of atomic particles. Specifically, a standing wave is comprised of a superposition of component waves traveling in two directions at the same time, with nodes corresponding to zero amplitude, while a trajectory involves motion in only one direction at a time, with no superposition. Furthermore, these eigenstates do not even describe quantum de Broglie waves of individual atoms of masses $M_1$ and $M_2$; rather, they describe a single quantum wave of a reduced mass $\mu$.

Can a set of discrete energy levels be compatible with a picture of quasi-classical trajectories? The discrete energy levels correspond to discrete oscillation amplitudes $A_n = (2E_n/K)^{0.5}$. Why are only certain oscillation amplitudes allowed, and all others apparently forbidden?

To address this question, first note that the quasi-classical free oscillation frequencies for the harmonic oscillator have the same value $\omega_0$ for any amplitude, so that a classical resonant interaction with an electromagnetic field at $\omega_0$ would be expected to either increase the amplitude of the oscillation, or to decrease it, depending on the relative phases of the two oscillations. These correspond to absorption or stimulated emission of radiation from the field. But the quantum properties of the electromagnetic field restrict energy exchange with the field to discrete multiples of $\hbar\omega_0$, corresponding to single photons. So if one starts with a molecule in the ground state, the only accessible excited states are those with energies $n\hbar\omega_0$ above the ground state.

In other words, transitions between quasi-classical collective modes oscillating at frequency $\omega_0$ are mediated by photons of the same frequency $\omega_0$. This may be expressed in terms of the Fourier transform $F(\omega)$ of the trajectory $\mathbf{r}(t)$, such that the transition rate to a state with higher (or lower) amplitude has the dependence

$$I(\omega) \propto |F(\omega_0)|^2 \tag{23}$$

This contrasts with a photon-mediated transition from one electron wave state to another (e.g., within an atom), whereby a quantum wave of frequency $\omega_1$ transitions to another quantum wave of frequency $\omega_2 = \omega_1 \pm \omega_0$.

In this quasi-classical picture, the photon energy $\hbar\omega_0$ delivered to the oscillator is actually shared (in inverse proportion to their masses) between the two atoms of the molecule. Therefore, while the total energy of the coupled system is quantized in units of $\hbar\omega_0$, their individual energies are not. In a multi-atom vibration (such as a phonon in a solid), the photon energy $\hbar\omega_0$ would be shared among a larger number of atoms.

Not all transitions of vibrational modes are produced by direct photon excitations. For example, one may have collisional excitation of vibrational modes. However, such a collisional excitation may still be electromagnetically mediated, and hence still subject to photon-based restrictions. On a microscopic level, all energy exchange may ultimately be attributable to photons, apart from those related to nuclear forces which may exhibit similar constraints. While this by no means proves the case, it does suggest a consistent mechanism for energy quantization in what is otherwise a set of classical trajectories.

Is there other evidence for quantum eigenstates in molecular vibrations apart from simple energy level differences? For example, vibrational wavefunctions (above the ground state) are standing waves with spatial nodes, which would be inconsistent with the present trajectory-based picture. But no direct experimental evidence for such nodes appears in the literature, suggesting that they do not exist.

One may apply a similar argument to the case of molecular rotation. Consider, for example, a diatomic molecule with masses $M_1$ and $M_2$. For rotations through an angle $\theta$, the classical Hamiltonian as a function of the angular momentum L is given by

$$H_c(L,\theta) = L^2/2I = L^2/2\mu R_0^2, \tag{24}$$

where $I = \mu R_0^2$ is the rotational moment of inertia. The solution yields the circular trajectory $\theta = \omega t = (L/\mu R_0^2)t$, where $\omega$ and $L$ can have any value. For the quantum case, a single-valued quantum phase of a matter wave with mass $\mu$ rotating around a loop gives rise to quantization of $L = n\hbar$, and energies $E_n = n(n+1)\hbar^2/2\mu R_0^2$. (This is the same problem as for an electron rotating around an atomic nucleus.) And indeed, infrared spectroscopy shows transitions between rotational states consistent with the quantum picture. But if there is not really a quantum wave corresponding to $\mu$, how can this be understood? Similarly to the vibration case, the same quantization of $L$ may arise for a quasi-classical trajectory, if one asserts that all transitions between quasi-classical rotational states require mediation of photons, each of which carries angular momentum $L_{ph} = \hbar$. Again, one might include collisional excitations in this, assuming that the collision involves an electromagnetic interaction.

Furthermore, this quantized angular momentum must be shared by the two (or more) atoms rotating about the common center of mass, so that while the angular momentum of the system is quantized, those of the individual atoms are not. Even within the conventional picture, one cannot obtain these rotational states from matter waves of the individual atoms.

It is useful to contrast the rotating molecule with the case of an electron in an excited state of orbital angular momentum. The P orbitals ($L=\hbar$) of electrons in molecules and crystals are normally present in standing-wave states, such as $P_x$, with angular nodes due to the superposition of degenerate rotational states in two directions about the same axis. This creates charge distributions that are essential for directional bonds. Analogous standing-wave rotational states for rotating atoms would be inconsistent with the trajectory-based picture, but again, these do not appear to be physically allowed states.

## V. Crystal diffraction and phonons

Diffraction is often viewed as the quintessential wave phenomenon.[20] It plays a central role in wave optics, and the observation of x-ray diffraction from crystals was critical historically in the early recognition that x-rays were in fact electromagnetic waves. If one regards a crystal as simply a periodic array of classical scatterers (Fig. 2a), then a diffraction pattern from an incident beam does indeed require that the beam consists of a phase-coherent extended wave.

In the conventional picture of coherent scattering of an extended wave with wave vector $\mathbf{k}_i$ from a crystal with a spatially varying scattering amplitude f(**r**), the Structure Factor F(**q**) is the spatial Fourier transform of f(**r**).[21] The diffracted wave intensity is then proportional to the square of F(**q**), where $\mathbf{q} = \mathbf{k}_f - \mathbf{k}_i$ is the change in the wave vector between the incident and scattered beams:

$$I(\mathbf{k}) \propto |F(\mathbf{q})|^2. \qquad (25)$$

If the scattering amplitude is periodic in space, then the structure factor exhibits peaks at the reciprocal lattice vectors **G**, and so does the coherently scattered intensity. This would apply equally well to a classical wave with wavevector $k_i$ and to a quantum wave with wave vector $\mathbf{k}_i = m\mathbf{v}/\hbar$.

However, a crystal is also a dynamically interacting set of atoms, with quantized vibrational excitations, i.e., phonons.[22] Within the present picture, these phonons are not quantum waves, but rather quasi-classical oscillating collective modes, with energies quantized in much the same way as for molecular vibrations described above. The existence of such phonons indicates that the crystal is not simply an ensemble of classical scatterers, and thus an alternative explanation for a diffraction pattern becomes possible. It is suggested here that just as a quasi-classical trajectory oscillating in time is restricted to energy changes $\Delta E = \hbar\omega$, a quasi-classical collective mode oscillating in space is restricted to momentum changes $\Delta \mathbf{p} = \hbar\mathbf{G}$. In this case, any entity that transmits momentum to the crystal, whether it is a particle or a wave, will scatter with its momentum change quantized according to this relation, thus producing the same diffraction pattern as for classical wave diffraction. Note the formal similarity between Eq. (25) and Eq. (23) above; in both cases, the transition rate or probability of the final state is proportional to the absolute square of the relevant Fourier transform.

Any crystal has a large number of discrete classical collective modes, each with frequency $\omega_n$ and corresponding wavevector $\mathbf{k}_n$. For a periodic lattice with reciprocal lattice vectors **G**, each mode may have spatial Fourier components at $\mathbf{k}_n+\mathbf{G}$ for all values of **G**. The vibrational amplitude of a given mode corresponds to energy and momentum quantized in multiples of $E_n = \hbar\omega_n$ and $\mathbf{p}_n = \hbar\mathbf{k}_n + \hbar\mathbf{G}$, corresponding to an integral number of phonons. Quantum transitions in a crystal typically correspond to creation or annihilation of a single phonon. In the present context, consider a **k**=0 phonon, where E=0 and **p**=$\hbar$**G**. This corresponds to the lattice as a whole absorbing a momentum $\hbar$**G**, exactly the values that correspond to standard lattice diffraction.

Within the particle trajectory picture (Fig. 2b), a small neutron strikes a single nucleus, being briefly absorbed into the nucleus before being re-emitted, either in the forward direction or in a different direction. If the direction is different, there is a momentum change $\Delta \mathbf{p}$, which must be absorbed either through recoil of the nucleus, or through movement of the crystal as a whole. Recoil of the nucleus would require generation of a phonon of non-zero energy, which would reduce the scattered energy of the neutron. Such inelastic scattering by neutrons is of course possible (and will be discussed further below), but there is a finite probability of elastic scattering, without net nuclear recoil. Because of the periodic lattice structure, and the fact that all transitions are subject to quantization rules, the only permitted transitions are ones that correspond to $\Delta \mathbf{p} = \hbar \mathbf{G}$ (including $\mathbf{G}=0$). What is quantized here is not the motion of the neutron, but rather the state of the crystal. So rather than having an extended quantum wave diffracting from a classical crystal, one has a quasi-classical particle inducing a transition in a quantized crystal. But remarkably, these two approaches produce essentially the same result – a crystal diffraction pattern.

The viewpoint of diffraction here may be somewhat similar to that by Van Vliet,[7,8] who derived momentum quantization of the crystal from conventional quantum formalism, but without explicit consideration of phonons. Furthermore, Van Vliet extended this analysis to a finite number of scatterers, with a similar result: a distributed wave is not necessary to obtain the conventional quantum diffraction.

This transition via emission of a phonon with $\mathbf{k}=\mathbf{G}$ and $\omega=0$ may be similar to what occurs in the Mössbauer effect.[23] A radioactive nucleus within a crystal emits an energetic gamma-ray photon, in many cases without nuclear recoil due to the absence of an emitted phonon. Because the frequency has not been Doppler-shifted, this same photon may be resonantly absorbed by a second nucleus, also without recoil. The present picture would suggest that the momentum of the Mössbauer photon should also be quantized to a value of $\hbar \mathbf{G}$ for the lattice. Since the photon wavevector for gamma radiation is typically much greater than that of the basis states of the reciprocal lattice, such momentum quantization should be obtainable.

Even if this quantized-lattice diffraction picture is correct, there are still situations where the conventional picture of coherent wave diffraction may be more appropriate. For example, one may have an electromagnetic wave with a large quantum number diffracting from a crystal lattice, where the wavefront is coherent and widely distributed. However, this suggests that diffraction due to incoherent electron or x-ray beams should perhaps be reexamined, to determine if this quantized-lattice picture may provide a more consistent picture in these cases as well.

Inelastic scattering in a lattice is generally considered as fundamentally different from elastic scattering that gives rise to diffraction. However, in the present picture they are quite closely related. The neutron would be briefly absorbed by a single nucleus, and the recoil would correspond to a phonon with $E=\hbar\omega$ and $\mathbf{p}= \hbar(\mathbf{k}+\mathbf{G})$. The intensity of scattering would be expected to go as the square of the Fourier transform of the dynamic response f(**r**,t) of a particular phonon mode (or other excitation):

$$I(\mathbf{q},\omega) \propto |F(\mathbf{q},\omega)|^2. \tag{26}$$

While the present analysis has focused on particle diffraction from periodic crystals, this may be directly generalized (as done by Van Vliet[8]) to other quantum diffraction effects in non-periodic structures, such as double-slit diffraction. In these cases, too, the quantum diffraction results of Eqs. (25) and (26) should be reproduced by consideration of momentum transfer from the slit structure without excitation of phonon-like modes. The same approach should also apply to the corresponding beam interference effects (as with neutron beams in a single-crystal interferometer[24]). It is notable that double-slit diffraction is often presented as a paradigm of the paradoxes of quantum measurement. Within the present picture, a small particle going through a slit without interacting with the edges would not be diffracted, but one that interacts with an edge can be subject to transverse momentum transfer, as determined by the Fourier transform of the scattering amplitude of the atomic distribution near the edges. While this result is not derived here, it would be expected to reproduce the conventional wave diffraction result within a picture of a scattering particle interacting with a quantized solid.

An alternative consistent explanation for quantum diffraction does not of course prove the absence of de Broglie waves of particles such as neutrons and atoms. But the evidence for such waves becomes weaker and more questionable. The next section addresses the phenomena of multi-particle quantum systems in more detail, including phenomena such as superfluid effects which would seem to require such coherent matter waves.

## VI. Composite particles and superfluids

Quantum theory identifies electrons (with spin ½) as fermions that follow the Pauli exclusion principle[25] limiting state occupation, and photons (with spin 1) as bosons that do not exhibit the Pauli limits. Conventional theory goes further to assert that combinations of electrons and other particles are themselves quantum waves that behave either as fermions or bosons, depending on the total spin of the composite. In some cases, an assembly of such composite bosons is found to exhibit a superfluid state, corresponding to a macroscopic quantum wave function. How can one understand these phenomena within the present picture, if such composite boson wave functions do not exist?

Consider a composite particle composed of two confined fundamental quantum waves, bound together. If these correspond to masses $M_1$ and $M_2$, each would exhibit its own oscillation with its own phase factor, e.g., $\phi_1 = k_1 x_1 - \omega_1 t_1$, and corresponding momentum $p_1 = \hbar k_1$ and $E_1 = \hbar \omega_1$ (including the relativistic rest energy $M_1 c^2$). Within the present picture, there is no quantum wave function of the combined particle that is different from those of its components.

In contrast, the conventional theory asserts that a "particle" composed of two quantum waves has its own wavefunction that is the product of the individual wavefunctions, even if the individual waves are non-interacting.

$$\Psi_{tot}(x_1,x_2,t_1,t_2) = \Psi_1(x_1,t_1)\, \Psi_2(x_2,t_2) \propto \exp[ik_1 x_1 - i\omega_1 t_1]\, \exp[ik_2 x_2 - i\omega_2 t_2] \quad (27)$$

If we assume now that the component waves oscillate in the same space and time so that $x_1 = x_2$ and $t_1 = t_2$, then one can directly combine the phase factors to obtain $\omega_{tot} = \omega_1 + \omega_2$ and $k_{tot} = k_1 + k_2$. This directly yields a de Broglie wave corresponding to $M_{tot} = M_1 + M_2$, provided that the constituent waves overlap in space and time. Further, it has the apparent advantage of enforcing conservation of energy and momentum. But note that a real wave oscillating at $\omega_{tot} = \omega_1 + \omega_2$ is physically quite different from two waves at $\omega_1$ and $\omega_2$. And if the nuclear wavefunction is restricted to the nucleus, then such a product wavefunction for an atom would also be restricted to the nucleus. This would not seem to be a logically reasonable representation of the quantum state of the entire atom.

There is no counterpart of wave multiplication in classical wave theory; one may add wave components, but one never multiplies them. This should be distinguished from wave effects in a nonlinear medium, where one does indeed generate sum and difference frequencies of all wave components. Similarly, an electromagnetic wave creates electric and magnetic potentials, which have the effect of effect of modulating the frequency and wavevector of a primary quantum wave such as an electron, again generating sum and difference frequencies. But this is quite different from simply multiplying two non-interacting waves.

Another role of wave function multiplication in conventional quantum theory is to account for the Pauli exclusion principle in pairs of identical fermions (such as electrons). Consider two electron states of the same spin with wavefunctions $\Psi_1$ and $\Psi_2$. One then constructs an antisymmetric linear combination of product wave function $\Psi_1\Psi_2$:

$$\Psi_{tot}(x_1,x_2,t_1,t_2) = \Psi_1(x_1,t_1)\,\Psi_2(x_2,t_2) - \Psi_2(x_1,t_1)\,\Psi_1(x_2,t_2) \qquad (28)$$

This combined wave function automatically goes to zero if the two wave functions $\Psi_1 = \Psi_2$, thus enforcing the Pauli exclusion principle for identical fermions. However, Eq. (28) also goes to zero identically for *any* wave functions, if the two oscillations are in the same spacetime (i.e., $x_1=x_2$ and $t_1=t_2$) as was done to obtain universal de Broglie waves below Eq. (27). It would seem illogical to assert that quantum oscillations should coexist in the same spacetime if they are different, but not if they are identical. Further, the physical significance of Eq. (28) is obscure; it would seem to represent a particular linear combination of two configurations with electrons interchanged, even if the two electrons are far apart. (This is not the same as a superposition of primary waves.) This provides a prototypical example of quantum entanglement,[10] which is intrinsically inconsistent with the concept of local realism of these physical systems. Eq. (28) can be easily extended to an arbitrary number of electrons through the use of Slater determinants,[26] but this quickly becomes truly complicated and even more heavily entangled. It is suggested in Section VII that this antisymmetric construction is artificial, and the Pauli principle can be more consistently regarded as a self-interaction of the primary electron field.

Another conventional concept that makes use of a composite quantum wave function is Bose-Einstein (BE) condensation.[27] This theory is based on the presence of a large number of identical de Broglie waves, heavily overlapping. In such a many-particle boson state, the particles (which may be atoms) are believed to condense into a state where most of the particles have the same quantum wave functions, and are oscillating in phase. This is analogous to the state of photons in a laser medium, for example. They also move together as a single coherent unit, forming a superfluid.[28] This is applied to superfluid helium-4 (He-4) at temperatures below 2.2 K, as well as to alkali gas atoms at ultra-low temperatures. This can only apply to bosons, which are integer-spin particles (like photons), as opposed to half-integer-spin particles (like electrons) which obey the Pauli exclusion principle, and hence cannot be strongly overlapping with the same quantum state. But in the present picture, an atom is *not* represented by a single quantum wave that includes both the nucleus and the electrons, and so should be neither a boson nor a fermion, and cannot be subject to BE condensation. Furthermore, the atom is not an extended overlapping wave, but rather a finite-sized particle – this is logically inconsistent with the physical basis for BE condensation.

A closely related phenomenon that is conventionally explained via a composite wave function is superconductivity, through the concept of a Cooper pair,[29,30] a bound electron-electron pair that is believed to function effectively as a boson, with a corresponding pair wavefunction. In the superconducting state, all such Cooper pairs have the same wavefunction, giving rise to a macroscopic quantum state with long-range order that is similar to BE condensation.

In fact, atoms interact with one another primarily via single-electron interactions, and the fermion nature of these electrons accounts for practically all of chemistry and condensed-matter physics. It is argued here that these superfluid properties may actually reflect a two-phase dense fermion packing of localized electron orbitals,[31] which yields long-range quantum coherence that is consistent with the Pauli exclusion principle. Within this picture, illustrated by the checkerboard pattern in Fig. 4a, each localized electron orbital is surrounded by several of the same electron orbitals in adjacent atoms, with the same energy and the same spin. By the Pauli principle, the adjacent coherent wavefunctions may not overlap; to prevent such overlap, they need to have a node between them. This can be obtained if there are two sublattices, with a phase difference of $\pi$ between their quantum phase factors. This is shown in two dimensions in Fig. 4a for simplicity (where the '+' and '-' represent the the two sublattices of the checkerboard), but a fully three-dimensional (3D) packing is envisioned. This is analogous to the two sublattices that are present in a 3D ionic crystal[32] such as NaCl, ZnS, or CsCl, with coordination numbers (numbers of nearest neighbors) between 4 and 8. For example, each Na atom is surrounded by 6 Cl atoms, and vice versa. Long-range structural order is not required here, so that the two-phase local correlated structure of an ionic liquid may be a better analogy. However, the packing of these two sublattices maintains long-range phase order over macroscopic distances, based entirely on fermion interactions. No boson interactions are required, and none are present.

Consider further the case of liquid helium, where each atom has two electrons of opposite spins. In the conventional theory, a He-4 nucleus (an alpha particle with 2 protons and two neutrons) with spin 0 is a boson, which has a wave function that combines with the two s-waves of the ground state electrons (also with net zero spin) to form a boson wave function for the atom as a whole. In contrast, a He-3 nucleus is a spin-1/2 fermion, which combines with the electrons to form a fermionic wave function for the atom as a whole. He-4 forms a superfluid below T=2.2 K, while He-3 does not form a superfluid until ~ 2 mK, 1000 times lower in temperature.[33] This is conventionally explained by Bose-Einstein condensation of He-4, versus no condensation for He-3 until magnetic nuclear interactions permit the formation of He-3 bound pairs (essentially the Cooper pairs of superconductivity) at much lower temperatures, which as bosons can then condense to the superfluid state.

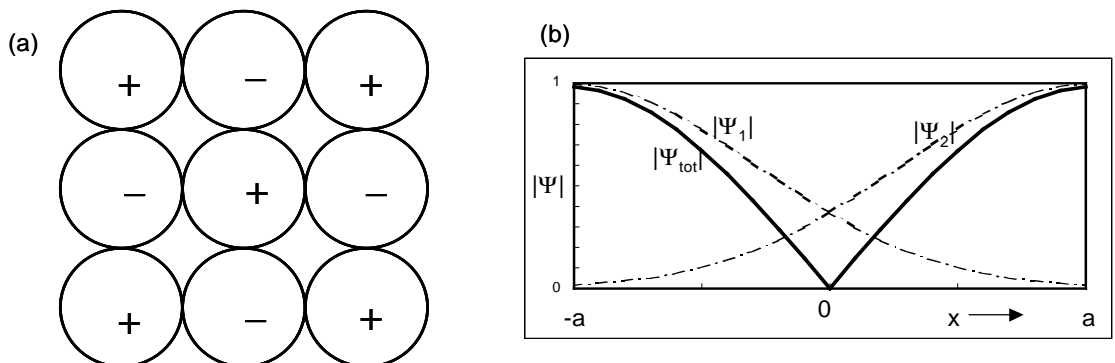

*Fig. 4. Superfluid formation via two-phase orthogonal fermion packing.*
*a) 2D checkerboard pattern illustrating two-phase packing.*
*b) Wavefunction dependence near inter-orbital node.*

In the present picture, there is no wave function for the helium atom, just those of the quarks in the nucleus (which are buried deeply in the center of the atom and are therefore largely irrelevant) and the electrons. So the interaction between adjacent helium atoms is dominated by direct-contact electron-electron interactions, as in conventional van der Waals interactions. Since the electron wavefunctions in He-4 and He-3 are very similar, how can one account for the dramatic difference in superfluid behavior?

But there is one significant difference between He-3 and He-4: He-4 has no unpaired spins, while He-3 has an unpaired spin in every atomic nucleus. Such spins do interact (weakly) with the electrons, and can induce spin-flip scattering of the electrons via spin exchange. In this regard, within the theory of superconductivity, it is well known that unpaired spins destroy superconducting order, even at low density.[34] In He-3, these unpaired spins are at very high density, and destroy what would otherwise be superfluid order. At much lower temperatures, these nuclear spins order (generally antiferromagnetically), suppressing spin-flip scattering. At that point, condensation into the superfluid state is again possible.

Looking more closely at the contact between adjacent atoms, consider spherically symmetric electron orbitals of the form $\Psi(r,t) = \psi(r)\exp(i\omega t)$, and for now consider only spin-up electrons. Take one atom centered at $x=-a$ and another at $x=+a$, and assume that the phase of the second atom is shifted by 180° ($\pi$ radians) from that of the first. Then the sum of these orbitals takes the form

$$|\psi_{tot}(x)| = |\psi(x-a)\exp(i\omega t) + \psi(a-x)\exp(i\omega t + \pi)| = |\psi(x-a) - \psi(a-x)| \qquad (29)$$

which indeed shows a node at $x=0$, half-way between the atoms, as shown in Fig. 4b. Note that this corresponds to an antisymmetric dependence of the total wavefunction, required by the Pauli principle, similar to that of Eq. (28), but based on addition of waves without any product states. A configuration of this sort is known in quantum chemistry as an anti-bonding state.[35]

Of course, each helium atom has 2 electrons, and a similar phase correlation would also be present for the spin-down electrons in the superfluid state. The spin-up and spin-down electrons are already orthogonal, and do not interact with each other, so their phases need not be correlated. In the presence of a uniform external magnetic field, the two spin states will have different energies (and hence frequencies), but the phase correlations for the orthogonal fermion packing will still be properly maintained for each spin state.

The situation in superconductors is somewhat more complicated, due to the fact that there are a large number of electron states of varying energies overlapping in the same region. A consistent real-space representation shows that electrons in a superconductor are localized orbitals (which however are not pinned to the crystalline lattice) with a size of order the superconducting coherence length $\xi(t)$ (typically ~ 100 nm in size), and there may be of order one million electron states in this region, all contributing to superconducting order.[31] What is needed, therefore, is the orthogonal packing (as in Fig. 4a) of these electron states of the same energy, with a similar packing for each discrete energy.

Electrons of different energy levels and spins are already orthogonal to each other; the 3D packing ensures that the electrons are also orthogonal within each energy level. There is no need for Cooper pairs of electrons to create a Bose-Einstein condensate. While these electron states have different energies, they all move coherently with the same velocity and a coherent spatially-dependent phase factor, exhibiting superconducting behavior. Furthermore, the factors of h/2e present in superconductivity theory, which are conventionally attributed to the 2e charge of a Cooper pair, can be attributed within the fermion packing picture to the interleaved two-phase nature of the ground state.[31,36]

Another type of superfluid consists of a dilute gas of alkali atoms at ultra-low temperatures, confined in a region using lasers and magnetic fields.[27] This is generally believed to be a true Bose-Einstein gas condensate, corresponding to atoms with a single valence electron and a nucleus with an odd number of nucleons, hence combining to form an atomic boson. While such atoms would normally freeze solid at these temperatures (due to bonding of atoms with opposite spin of the valence electron), the magnetic field causes the spins of the electrons to align, eliminating the bonding energy and suppressing solidification. At sufficiently low temperatures (in the nK range), this dilute gas exhibits coherent effects indicative of superfluidity and quantum coherence.

Can these results be understood within the picture of Fig. 4a, in the absence of boson formation? The orthogonal fermion packing of Fig. 4a requires dense correlated packing of a liquid or a solid, and is not compatible with a dilute gas. It is suggested here that the superfluid behavior of these alkali atoms may be associated with the formation of nanoscale droplets, i.e., clusters of atoms, within which the correlated structure of Fig. 4a occurs. But this needs to be analyzed further.

There have also been observations of superfluid condensation in alkali atoms that are believed to be fermions,[37] rather than bosons. This fermion condensation is conventionally attributed to the formation of Cooper pairs of alkali atoms, which can then condense as bosons. However, these may also be due to droplet formation and ordering within the orthogonal fermion packing picture of Fig. 4a.

Finally, is there other evidence for de Broglie waves in composite objects such as atoms and molecules? Diffraction effects with atoms and even molecules have been reported,[38] but these can be attributed to quantized momentum transfer as for the case of neutrons in Section III. Another observed effect is quantization of circulation in superfluids,[39] whereby the angular momentum around a loop is quantized in units of $\hbar$. This is conventionally interpreted as being due to a single-valued de Broglie wave around the loop, where $\lambda=h/mv$ and m is the atomic mass, but it may alternatively be explained by quantized rotational transitions of the macroscopic superfluid, as suggested for quantized rotational states of molecules in Section IV.

Taken together, the conventional evidence for de Broglie waves of composite particles seems to be weak, and can be alternatively explained by the present picture. Furthermore, multiplication of wavefunctions appears to be an unnecessary mathematical artifact that leads to non-physical results.

## VII. Basis for primary wave quantization

The present paper focuses on how classical and quantum behavior of composites may derive from the fundamental quantum waves, rather than on the origin of these quantum waves. However, this Section provides a brief outline showing that electrons and photons may constitute real-space relativistic waves with rotating vector fields (much like classical electromagnetic fields) that carry distributed spin angular momentum, and that this spin is quantized in units of $\hbar$ or $\hbar/2$. It is argued that this spin quantization reflects spontaneous symmetry-breaking that stems from a nonlinear self-interaction that is not evident in the Schrödinger equation. Such a system is compatible with the models presented in this paper, and also maintains local realism without entangled states.

First, let us show that the solution to the Schrödinger equation represents a real relativistic wave, even in its conventional form as a non-relativistic equation for a complex wavefunction. The key equations are the relations made famous by Einstein, $E=mc^2$ and $E=hf$.

Consider a real oscillating field $F(x,t)$, and assume for the moment that this is a scalar field. Assume further that this field oscillates coherently and uniformly in a given region of space: $F(x,t) = F_0 \cos(\omega_0 t)$. This represents a wave with $k=0$ and hence has an infinite wavelength and an infinite phase velocity $v_{ph} = \omega/k$. However, its group velocity is $v_g = \partial\omega/\partial k = 0$; this is the rest frame for this wave. If one changes to another reference frame moving with velocity v, the wave is subject to a relativistic Doppler shift,[40] which shifts both the frequency and the wavelength. Spatially offset oscillations that are simultaneous in one reference frame will not be simultaneous in another reference frame; this is a unique characteristic of relativity. In contrast, a classical non-relativistic Doppler shift will change the frequency, but the wavelength must stay the same; time is absolute, so events that are simultaneous in one reference frame must be simultaneous in all reference frames.

Application of the Lorentz transform between references frames yields the results:

$$\omega' = \gamma\omega_0 \quad \text{and} \quad k' = \gamma v\omega_0/c^2, \tag{30}$$

or equivalently one has a wave dispersion relation

$$k'^2 c^2 = \omega'^2 - \omega_0^2, \tag{31}$$

where $\gamma = (1-v^2/c^2)^{-1/2}$ is the standard factor in special relativity. The phase velocity (dropping the primes) is $v_{ph} = \omega/k = c^2/v > c$, but the physical speed of the wave (and a wave packet constructed from it) is the group velocity $v_g = \partial\omega/\partial k = kc^2/\omega = v$, as is required for consistency.

Note that $\omega$ and $k$ transform as a relativistic 4-vector in the same way as E and p, where $\omega_0$ is analogous to the rest energy $mc^2$. These relations are valid for any classical relativistic wave with a rest frame, and are not special for quantum waves. If one makes the additional identification that $\omega_0 = mc^2/\hbar$, this immediately corresponds to a de Broglie wave with $k = p/\hbar$. The only quantum aspect here is that the total energy of this wave is quantized. This was first derived by de Broglie[41] early in the development of quantum mechanics, but the significance of this is often forgotten. The de Broglie wave makes sense only as a true relativistic wave – it has no non-relativistic limit.

The dispersion relation Eq. (31) corresponds to the Klein-Gordon differential wave equation[42]

$$\partial^2 F/\partial t^2 = c^2 \nabla^2 F - (mc^2/\hbar)^2 F \qquad (32)$$

More generally, if this quantum wave is subject to a spatially dependent potential energy $V(r)$, the rest energy and the rest mass shift appropriately, as does the frequency of the de Broglie wave:

$$\hbar\omega_0 = mc^2 = m_0 c^2 + V(r). \qquad (33)$$

(Of course, this requires that the zero of potential energy is consistently defined.) Eq. (31) then becomes:

$$\partial^2 F/\partial t^2 = c^2 \nabla^2 F - (m_0 c^2/\hbar + V(r)/\hbar)^2 F \qquad (34)$$

Note that this is a real equation for a real oscillating wave; F is not a complex wavefunction. But let us define a new complex wavefunction $\Psi$ that represents the wavevector F shifted down in frequency by $\omega_0$:

$$F = \Psi \exp(-i\omega_0 t) \qquad (35)$$

Substituting this into Eq. (34) yields

$$\partial^2 \Psi/\partial t^2 - 2i\omega_0 \partial\Psi/\partial t - \omega_0^2 \Psi = c^2 \nabla^2 \Psi - \omega_0^2 \Psi - (2\omega_0 V/\hbar) \Psi - (V^2/\hbar^2) \Psi. \qquad (36)$$

In the non-relativistic limit, F corresponds to a wave with a narrow frequency bandwidth around $\omega_0$, so that the relevant frequency components of $\Psi$ after frequency shifting are $\ll \omega_0$. In that case, the first term on the left side of Eq. (36) is a small higher order term (compared to the second and third terms) and can be dropped. The term in $V^2$ on the right is another small higher order term for the usual case that $V \ll m_0 c^2$, and it too can be dropped. The remaining terms lead directly to the usual time-dependent Schrödinger equation.

$$i\hbar \partial\Psi/\partial t = (-\hbar^2/2m) \nabla^2 \Psi + V(r) \Psi \qquad (37)$$

Note that the speed of light c no longer appears in Eq. (37), hiding its relativistic origin. But the fact that the wavevector k depends on velocity (as in the de Broglie relation) can only be understood as consequence of relativity. It may seem odd that one has relativistic effects for a case where $v \ll c$, but the real physical wave is not $\Psi$ but rather F, which is oscillating at the enormous frequency $m_0 c^2/h$ (~ $10^{20}$ Hz for an electron). Extremely small time shifts, as would arise from special relativity for small velocities, can correspond to large phase shifts.

This derivation assumed a scalar field F, but the argument also holds for a vector field **F**. One aspect of a real oscillating vector field is its polarization, and one type is circular polarization (CP) associated with a rotating field of fixed magnitude. It is argued here that *all* fundamental quantum waves are really localized vector fields rotating about a spin axis, where the quantum phase factor is the rotation angle.[43] Further, the total angular momentum about this axis (which is Lorentz invariant) represents the spin of the quantum wave, which is quantized to $\hbar/2$ for an electron or similar fundamental fermion, and $\hbar$ for a photon or similar boson.

One can motivative this general argument by consideration of a CP classical electromagnetic wave,[44] which is a transverse wave with the **E** and **H** fields rotating in the plane perpendicular to the direction of wave motion. It is well known that an electromagnetic wave carries distributed energy and momentum, associated with the Poynting vector $\mathbf{P} = \mathbf{E} \times \mathbf{H} = (\mathbf{E} \times \mathbf{B})/\mu_0$, and distributed through the wave. Here **B** and **H** are the usual magnetic vectors (SI units are used here and throughout the paper). One can define an energy density $\mathcal{E}$ and momentum density $\mathcal{P}$ given by the following expressions:

$$\mathcal{E} = |\mathbf{P}|/c = |\mathbf{E} \times \mathbf{B}|/\mu_0 c = \varepsilon_0 \mathbf{E}^2 \tag{38}$$

$$\mathcal{P} = \mathcal{E}/c = (\mathbf{E} \times \mathbf{B})/\mu_0 c^2 = \varepsilon_0 \mathbf{E}^2/c \tag{39}$$

It is somewhat less well known (but still a standard result[45]) that for a CP wave, one may also define a spin angular momentum density $\mathcal{S}$, which is given by

$$\mathcal{S} = |\mathbf{E} \times \mathbf{A}|/\mu_0 c^2 = \varepsilon_0 \mathbf{E}^2/\omega, \tag{40}$$

where **A** is the usual vector potential given by $\mathbf{B} = \nabla \times \mathbf{A}$, and we are assuming a monochromatic wave of frequency $\omega$. This spin density $\mathcal{S}$ therefore has the following relations to $\mathcal{E}$ and $\mathcal{P}$:

$$\mathcal{E} = \mathcal{S}\,\omega \quad \text{and} \quad \mathcal{P} = \mathcal{S}\,k, \tag{41}$$

where as usual for an EM wave in free space, $k = \omega/c$. Now let us further assume that a one-photon state consists of such a CP wave packet (or otherwise confined wave), where the total angular momentum, integrated over the volume of the localized wave, is quantized to the value $S = \hbar$, which is indeed accepted as the spin of a single photon. Then the standard Einstein-de Broglie relations follow directly by integrating Eq. (41) over the same volume:

$$E = \hbar\omega \quad \text{and} \quad p = \hbar k. \tag{42}$$

One can construct a similar argument for an electron, if one assumes that in this case, $\mathcal{S} = 2\mathcal{E}/\omega$. Then quantization of spin with S= $\hbar/2$ again yields the standard Einstein-de Broglie relations. Unlike the case of the photon, one can transform to the rest frame of an electron, so that the spin axis is independent of the direct of electron motion. Note that angular momentum is a Lorentz invariant, so that the spin is $\hbar/2$ in any reference frame.

Of all of the particles in the standard model of particle physics, almost all are either spin-½ fermions like the electron (neutrinos, quarks, etc.), or spin-1 bosons like the photon (gluons, etc.). The only exception is the Higgs boson,[46] which is believed to be a scalar particle with zero spin. Such a scalar particle would seem to be in conflict with the present physical picture, with its central role for rotating vector fields with quantized total spin. But at least as of 2011, the Higgs Boson has not yet been observed; perhaps *the Higgs Boson does not exist*, and an alternative mechanism is responsible for the properties of the weak interaction and of masses within the standard model.

These observations suggest that there is a special role for spin in quantum mechanics. One may have electron states which are superpositions of components with different values of orbital angular momentum (for example, the $P_x$ orbital discussed earlier), but the spin angular momentum is always $\hbar/2$ with a specific spin axis. Similarly, a single photon is always CP with spin $\hbar$, although one can construct multi-photon states with other values of spin. For example, within this picture, one cannot have a linearly polarized single photon, although a two-photon field with linear polarization can be constructed as the sum of two correlated CP photons with opposite helicity. This may be relevant to some of the measurement paradoxes discussed in the next section.

The argument thus far indicates that *if* spin is quantized, then the other equations of quantum mechanics automatically follow. However, this does not explain why spin should be quantized in what is otherwise a classical relativistic field, or provide a mechanism to do this. Indeed, the Schrödinger equation (and the relativistic Klein Gordon equation) is an equation for the evolution of a wave in a potential, and as a linear equation would be valid for any amplitude of the wave, and therefore for any spin. Something else which is not included in these equations must be quantizing the spin.

One suggested mechanism for spin quantization[47] might be called "Spontaneous Quantum Domain Formation", in analogy with formation of magnetic domains in a ferromagnet. In the ferromagnetic case, the presence of local exchange interactions between adjacent magnetic atoms spontaneously leads to the formation of an array of macroscopic magnetic domains, where each domain acts as if it is a macroscopic magnetic particle with a proportionally large magnetic moment. The evidence for the microscopic exchange interaction is largely hidden, except for brief time intervals during which the configuration of the domains changes. In the quantum case, one would similarly have a hidden local self-interaction in the quantum field that spontaneously gives rise to quantization of spin within each coherent quantum domain. Such a self-interaction would presumably be highly nonlinear (e.g., self-modulation), and would not appear in the standard linear wave equations such as the Schrödinger equation. This self-interaction would become dominant only during transitions between quantum states, which amounts to a reconfiguration of these quantum domains.  Such a reconfiguration would be microscopically mechanistic rather than probabilistic, and consistent with special relativity rather than instantaneous or non-local.  It could even shift fractional spin components among multiple quantum domains, so long as the initial and final states maintained proper spin quantization.   The indistinguishability of identical quantum particles is a natural consequence, without requiring any product states or entanglement.  Furthermore, this self-interaction would also be responsible for the Paul exclusion principle, which is really an interaction rather than an accounting rule.  Finally, domain formation is an example of symmetry breaking (as is the Higgs field), and the mechanism for quantum domain formation might also be responsible for fundamental masses of quantum particles. This picture of spontaneous quantum domain formation is certainly speculative, and beyond the purview of the present analysis, but it provides a plausible outline of a way to achieve quantization from otherwise continuous fields, that may be fully compatible with both special relativity and local reality.

The argument presented here suggests that both fundamental fermions and bosons have intrinsic quantization of spin.  A possible logical alternative might be that only the fermions are directly quantized; the bosons could appear quantized because of transition rules between fermion states.  But this would lead to the proliferation of fractional "pieces of photons", creating a background noise spectrum.  This may be similar to a different proposed alternative to standard quantum theory known as "stochastic electrodynamics".[48]

**VIII. Quantum Measurement and Quantum Information**

The field of quantum measurement has a complicated history of both theory and experiments, together with a set of paradoxes. It is generally asserted that the accepted results of quantum mechanics are inconsistent with any local realistic picture, due largely to the presence of quantum entanglement. It is suggested here that these conclusions may be based on several misunderstandings as to the physical basis of quantum states, and the experiments may not really be proving what is asserted. Furthermore, in recent years the new field of quantum information has developed, including both quantum computing and quantum communication, built around quantum entanglement as a fundamental assumption. While a complete analysis of these issues is beyond the scope of the present paper, several key points are briefly discussed below.

First, consider the conventional theory of quantum measurement,[49] which is based on the postulate of projection. Essentially, a quantum state evolves as a superposition of basis states in abstract Hilbert space until a measurement (an interaction with a classical observer or instrument) forces it into one of the states, with a probability amplitude based on the projection of the state vector onto the measurement vector. This projection involves an instantaneous "collapse of the wave function", which would seem contrary to the spirit of special relativity. Given the degree to which quantum mechanics is fundamentally based on special relativity (as shown in Section VII), this sudden collapse would seem to be unlikely. In contrast, the present picture suggests all measurements are transitions subject to quantum selection rules, with real-time dynamics consistent with local realism. For example, consider an extended photon wave which may excite any one of a large number of atoms in a detector. Which atom is ultimately excited depends on a set of uncontrolled initial conditions, including relative phases of the photon and the various electron wavefunctions. The transition is a continuous dynamical process, which may be fast but not instantaneous.

A central paradox of quantum measurement is the Schrödinger cat paradox.[50] This was intially introduced by Schrödinger around 1950 as an indication that quantum theory was incomplete and inconsistent, but it is now conventionally accepted, at least for an inanimate object in a coherent quantum state. In one form of this paradox, it is asserted that a Schrödinger "cat" may be in a quantum state that is a linear superposition of being alive and being dead, depending on the coupled state of a radioactive atom. It is not until the cat interacts with an appropriate classical observer that the coherent quantum state is projected into one of the two basis states. In contrast, within the present picture, only fundamental quantum fields such as electrons are subject to linear superposition, so that a superposition of a live cat state and a dead cat state makes no sense at all. Furthermore, the entangled coupling of a microscopic quantum state with a macroscopic object also makes no sense.

Further paradoxes are associated with measurement of correlated quantum states of two or more quantum objects, in particular the EPR paradox, named after a paper by Einstein, Podolsky, and Rosen.[51] Although there are various versions of this paradox, the general argument is that two correlated quantum "particles" are initially prepared such that their combined value of a given property (e.g., momentum or spin) is well defined, but their corresponding individual values are not well defined until a quantum measurement is made on one of the particles. Then this first wave function collapses, but since this is a coupled entangled state of the two particles, the wave function of the second particle must also collapse at the same time, in order to maintain a complementary value to that of the first particle. Since the two particles can be far apart at the time the first measurement is made, this would seem to violate (at least the spirit of) special relativity. Alternatively, if no such instantaneous collapse occurs, then each of the two particles must remember their initial preparation in a way that incorporates the final result, via some sort of "hidden variables". But a general analysis of these types of correlated measurements by Bell has led to a set of inequalities that constrain the existence of "local hidden variables".[52] A number of experiments have been done, practically all using correlated photons and polarization measurements, and these tend to confirm the standard quantum predictions, as opposed to an alternative explanation based on local hidden variables.[53] These results have been generally interpreted to rule out any alternative to standard quantum mechanics, although some questions about possible "loopholes" in the results continue to be discussed.[54]

The present picture questions the real existence of the entangled product states that are used in the conventional explanation of these EPR-type experiments, and instead proposes that each quantum wave represents a localized real-space rotating vector field consistent with local realism. Furthermore, in terms of the optical experiments, the present picture suggests that single photons are necessarily circularly polarized with spin $\hbar$, in contrast to linearly polarized single photons with zero spin, which are essential to the interpretation of many of these measurements. From this point of view, an electromagnetic wave that passes through a linear polarizer must be a superposition of at least 2 counter-rotating CP photons. It would be interesting to re-analyze the results of these experiments with this picture in mind, to see if this could account for the measured results in a way that does not require non-locality.

The field of quantum information also depends on entangled states of multiple quantum systems, using the generalized Hilbert space approach. A qubit[55] or quantum bit is a quantum state that can hold a linear combination of two values. If a large number of qubits are coupled together, the resulting multiply entangled state enables massive parallelism in computing, which cannot be achieved using classical computing approaches. This has been proposed not only for microscopic quantum states such as individual electrons or atoms, but also for macroscopic systems such as superconducting circuits.

In contrast, the present picture questions the existence of entanglement among primary quantum waves, and furthermore suggests that macroscopic systems are not really quantum waves at all, even if their energy levels are quantized. If true, the entire basis for most quantum computing is questionable. But in any case, the realistic picture of quantum waves presented in this paper should lead to a re-examination of the physical basis of quantum computing and quantum communication.

## IX. Conclusions

Quantum mechanics has been full of inconsistencies and paradoxes from the very beginning. The intervening century has led *not* to increased clarity, but rather to abstract formalisms that are largely divorced from physical pictures and physical reality. Many of the early pioneers of quantum theory, such as Einstein, de Broglie, and Schrödinger, were quite uncomfortable with the direction that the theory has taken. The present paper incorporates a radically different approach, based on real-space physical pictures and special relativity, in a way that might have appealed to some of these pioneers.

This requires a sharp departure from conventionally accepted physical pictures. It is suggested here that one may better understand the basis for the microworld within a hierarchical picture, where true quantum waves are only present at the bottom of the hierarchy (the electrons, photons, and quarks of the standard model), and confined quantum waves generate classical physics. This is not merely speculation; it derives simply from the standard Schrödinger wave equation. Quantum waves are not universal aspects of all particles, but rather provide a way to quantize the primary fields. There are no composite quantum waves, and no multi-particle entangled states. Diffraction can be viewed as a particle phenomenon involving quantized momentum transfer. Superfluid behavior is due not to Bose-Einstein condensation of composite bosons, but rather to a two-phase fermion correlation of valence electrons. Quantum measurement may be better viewed as quantized transitions, mediated by a local dynamical process, rather than a sudden wavefunction collapse. And the quantum computing paradigm may be based on fundamental misunderstandings of the nature of quantum waves.

These are rather provocative assertions, contradicting much that is universally accepted among generations of physicists. But this represents the logical implications of a simple, consistent picture of reality. This paper is intended to start a new discussion about these issues, which are generally discouraged or pushed to the fringes. Only with such an open discussion can progress in this field be made.